\documentclass[floats,prd,showpacs,preprint]{revtex4}
\usepackage{latexsym}
\usepackage{amsmath}
\usepackage{amssymb}
\usepackage{amsfonts}
\usepackage{graphicx}
\usepackage{dcolumn}

\begin{document}

\title{Renormalization Group Approach to Causal Bulk Viscous Cosmological Models}

\author{J. A. Belinch\'on}
\email{abelcal@ciccp.es}
\affiliation{\small {Grupo Inter-Universitario de An\'{a}lisis Dimensional.\\ Dept. F\'{\i}sica ETS Arquitectura UPM Av. Juan de Herrera 4 Madrid 28224 Espa\~na}}
\author{T. Harko } 
\email{tcharko@hkusua.hku.hk}
\affiliation{Department of Physics, The University of Hong Kong,
Pokfulam Road, Hong Kong, P. R. China.}
\author{M. K. Mak}
\email{mkmak@vtc.edu.hk}
\affiliation{Department of Physics, The Hong Kong University of Science and Technology,Clear Water Bay, Hong Kong, P. R. China.}

\date{March 26, 2002}

\begin{abstract}
The renormalization group method is applied to the
study of homogeneous and flat Friedmann-Robertson-Walker type Universes, filled with a
causal bulk viscous cosmological fluid. The starting point of the study is the consideration of the scaling
properties of the gravitational field equations, of the causal evolution equation of the bulk
viscous pressure and of the equations of state. The requirement of
scale invariance imposes strong constraints on
the temporal evolution of the bulk viscosity coefficient, temperature and relaxation time, thus
leading to the possibility of obtaining the bulk viscosity coefficient-energy density dependence. 
For a cosmological model with bulk viscosity coefficient proportional to the Hubble parameter,
we perform the analysis of the renormalization group flow around
the scale invariant fixed point, therefore obtaining the long time behavior of
the scale factor. 
 
\end{abstract} 

\pacs{98.80.Hw, 98.80.Bp, 04.20.Jb}

\maketitle

\section{Introduction}

The renormalization group (RG) has been proved to be a powerful tool in
statistical physics, quantum field theory, critical phenomena and chaotic
dynamic \cite{WiKo74,Sa98}. The essence of the renormalization group method
is to extract structurally stable features of a system, which are
insensitive to details. Many physical systems in field theories, polymers
etc. exhibit universal scaling functions and critical exponents in the limit 
$\Lambda/\xi$, where $\Lambda$ is some ultraviolet cut-off and $\xi$ is a
temperature-dependent correlation length. The RG group method is the main
tool with which to elucidate this universal behavior. On the other hand, from
a purely mathematical point of view, the RG is regarded as a means of
asymptotic analysis. Long-time asymptotics of non-linear partial
differential equations using RG ideas have been studied recently by
Bricmont, Kupiainen and Lin \cite{BrKuLi94} and by Chen, Goldenfeld and Oono 
\cite{ChGoOo96}. In this formulation, the RG transformation is the
integration of the equation up to a finite time, followed by a rescaling of
the dependent and independent variables. Therefore the problem at infinite time is
reduced to the problem at finite time. A fixed point of the RG
transformation corresponds to a scale-invariant solution of the differential
equation, and the long-time behavior of the solution can be obtained by
studying the flows around fixed points. The RG method for reduction of
evolution equations has been formulated in terms of the notion of invariant
manifolds in \cite{EiFuKu00}, and it has been shown that the perturbative RG
method constructs invariant manifolds successively, as the initial value of
the evolution equation.

The RG method has been also intensively applied in the study of cosmological
models and of the critical behavior of the gravitational collapse of a
radiation fluid. Koike, Hara and Adachi \cite{KoHaAd95} presented a scenario
based on the renormalization group that can explain the universality and
scaling observed in a numerical study of gravitational collapse of radiation
fluid. This approach for understanding and analyzing critical behavior in
the gravitational collapse has been generalized, to a general framework, for
perfect fluids with pressure proportional to density and with arbitrary $%
\gamma$ \cite{HaKoAd96}. Critical phenomena occur for $1<\gamma<1.88$ and
the uniqueness of the relevant mode around a fixed point is established by
Liapunov analysis.

The averaging hypothesis tacitly assumed in standard cosmology was discussed
by using RG methods by Carfora and Piotrkowska \cite{CaPi95}. A critical
behavior, which can be related to the formation of sheet-like structures in
the Universe, is obtained, and explicit expressions for the normalized Hubble
constant and for the scale dependence of matter distribution are also
provided. The observable scaling properties in Newtonian cosmology have been
considered in \cite{SaKoMaKuMoNa98}. The quantitative scaling properties
have been analyzed by calculating the characteristic exponents around each
fixed point, thus leading to the possibility of a fractal structure of the
Universe beyond a horizon scale. The RG method has been applied to the
Einstein equations in cosmology by Iguchi, Hosoya and Koike \cite{IgHoKo98},
who carried out a detailed analysis of renormalization group flows in the
vicinity of the scale invariant fixed point in the spherically symmetric and
inhomogeneous dust filled Universe. The first order solution of
long-wavelength expansion of the Einstein equations has been improved in 
\cite{NaYa99}. By assuming that the RG transformations have the property of
Lie group, the secular divergence caused by the spatial divergence terms can
be regularized and absorbed to the background seed metric. The renormalized
metric describes qualitatively quite well the behavior of the gravitational
collapse in the expanding Universe. The back reaction effect due to the inhomogeneities
has been incorporated into the framework of the cosmological perturbations, using the RG method,
by Nambu \cite{Na00}. The RG equations, which determines the dynamics of the space-time, have been
derived in a gauge invariant manner.

The basic techniques and ideas for the construction of RG theories for the study
of non-equilibrium processes in inflationary and semiclassical  cosmology and
stochastic gravity have been recently reviewed in \cite{CaHuMa01}. The complete study of the
flow of the renormalization group in the Einstein-Hilbert truncation of quantum gravity in
four dimensions and the classification of the RG trajectories have been done in
\cite{ReSa02}. Cosmologies with a time-dependent Newton constant and a cosmological constant
have been investigated, from the RG point of view, by Bonnano and Reuter \cite{BoRe02}. The scale
dependence of $G$ and $\Lambda $ is governed by a set of RG equations, which are coupled to the
Einstein equations in a consistent way. This model could explain the data from recent
observations of red-shift of Type Ia Supernovae without introducing a quintessence field.
The cosmological implications of the running of the cosmological constant and whether the
scaling dependences of $G$ and $\Lambda $ spoil primordial nucleosynthesis have been discussed, using RG techniques, in \cite{ShSo02}.

It is the purpose of the present paper to apply the RG group
techniques to the analysis of the bulk viscous cosmological models. 

Dissipative processes are supposed to play a fundamental role in the
evolution and dynamics of the early Universe. Over thirty years ago Misner \cite
{Mi66} suggested that the observed large scale isotropy of the Universe is
due to the action of the neutrino viscosity, which was effective when the
Universe was about one second old. Bulk viscosity may arise in different
contexts during the evolution of the early Universe. Some physical processes
involving viscous effects are the evolution of cosmic strings, the classical
description of the (quantum) particle production, interaction between matter
and radiation, quark and gluon plasma, interaction between different
components of dark matter etc. \cite{ChJa96}.

The first attempts at creating a theory of relativistic dissipative fluids
were those of Eckart \cite{Ec40} and Landau and Lifshitz \cite{LaLi87}.
These theories are now known to be pathological in several respects.
Regardless of the choice of equation of state, all equilibrium states in
these theories are unstable and in addition signals may be propagated
through the fluid at velocities exceeding the speed of light \cite{Is76}.
The problems arise due to the first order nature of the theory, i.e., it
considers only first-order deviations from the equilibrium leading to
parabolic differential equations, and hence to infinite speeds of propagation
for heat flow and viscosity, in contradiction with the principle of
causality. Infinite propagation speeds already constitute a difficulty at
the classical level, since one does not expect thermal disturbances to be
carried faster than some (suitably defined) mean molecular speed.
Conventional theory is thus applicable only to phenomena which are
quasi-stationary, i.e., slowly varying on space and time scales characterized
by mean free path and mean collision time \cite{IsSt76}. This is inadequate
for many situations in high-energy astrophysics and relativistic cosmology
involving steep gradients or rapid variations. The deficiencies can be
traced to the fact that the conventional theories (both classical and
relativistic) make overly restrictive hypothesis concerning the relation
between the fluxes and densities of entropy, energy and particle number.

A relativistic second-order theory was found by Israel \cite{Is76} and
developed by Israel and Stewart \cite{IsSt76}, into what is called transient
or extended irreversible thermodynamics. In the second order theory the
deviations from equilibrium (bulk stress, heat flow and shear stress) are
treated as independent dynamical variables leading to a total of 14
dynamical fluid variables to be determined. However, Hiscock and Lindblom 
\cite{HiLi89} and Hiscock and Salmonson \cite{HiSa91} have shown that most
versions of the causal second-order theories (the so-called truncated models)
omit certain divergence terms. When these terms are added to the evolution equation
of the bulk viscous pressure, one obtain the full causal theory, with well behaved solutions
for all times. Causal bulk viscous thermodynamics has been
extensively used for describing the dynamics and evolution of the early
Universe and some astrophysical process \cite{HiLi87,Ma95}.

The fact that the Universe has some hierarchical structure, and that the
scale invariant solutions of the gravitational field equations plays an
important role is well known \cite{IgHoKo98}. The homogeneous Einstein
equations are scale invariant. Using this property, we apply the RG
method to study the long-time asymptotics of the general relativistic field
equations, describing the evolution of a homogeneous causal bulk viscous
fluid filled Universe. For the RG method we closely follow the approach
presented in \cite{IgHoKo98}. 

The renormalization group approach shows the particular role played by the
exponent $s$ of the density in the equation of state of the bulk viscous
coefficient. The gravitational field equations are invariant under scale
transformations only for the value $s=1/2$ of the exponent.
Hence this method provides an effective and efficient tool for
investigating general properties of cosmological models.

The present paper is organized as follows. In Section II we give a brief and formal
presentation of the renormalization group method and its applications to the study
of the asymptotics of differential equations.
The scaling properties of the basic equations of the causal thermodynamics in a
homogeneous space-times are investigated in Section III, and their implications
on the equation of state of the cosmological causal bulk viscous fluid are considered.
The flow in the RG around the fixed points of the renormalization group equations is analyzed in Section
IV. Finally, in Section V we discuss and conclude our results.

\section{Renormalization group and differential equations}

In this Section we briefly review, from a formal and abstract mathematical
point of view, the formalism of renormalization group, as applied to the
study of systems of ordinary or partial differential equations.
The presentation closely follows \cite{BrKuLi94} and \cite{HaKoAd96}, and for further
details and specific applications the reader is referred to these papers.

The basic idea in the RG formalism is to consider the time evolution of initial data as
a flow of a renormalization group transformation. Suppose we are interseted
in the time evolution of $n$ unknown functions $y=(y_{1},y_{2},...,y_{n})$,
which are real-valued functions of the time $t$ and of a spatial coordinate $%
x$, and satisfy a system of partial differential equations
\begin{equation}
F\left( y,\frac{\partial y}{\partial t},\frac{\partial y}{\partial x}%
,t,x\right) =0.  \label{ev1}
\end{equation}

The time evolution of the system (\ref{ev1}) is determined once one
specifies the values $y\left( t,\cdot \right) $ at the initial time $t=1$.
The space of functions $y\left( t,\cdot \right) $ at a fixed time is called
the phase space $\Gamma $ of the system, and assume it is a real vector
space. As a first step in introducing the renormalization group
transformation, we define the scaling transformation $S\left( L,\vec{\alpha}%
,\beta \right) $, which depends on the real parameters $L\in R,\vec{\alpha}%
=\left( \alpha _{1},\alpha _{2},...,\alpha _{n}\right) \in R,\beta \in R$ \cite{HaKoAd96}:
\begin{equation}
S\left( s,\vec{\alpha},\beta \right) :y_{i}\left( t,x\right) \rightarrow
y_{i}^{\left( L,\vec{\alpha},\beta \right) }\left( t,x\right) \equiv
e^{\alpha _{i}L}y_{i}\left( e^{-L}t,e^{-\beta s}x\right) .  \label{scaltr}
\end{equation}

If $y$ is a solution of the Eq. (\ref{ev1}), and if with a suitable choice
of the constants $\vec{\alpha},\beta $ and for an arbitrary scale parameter $%
L$ (close to zero), the new function $y^{\left( L,\vec{\alpha},\beta \right)
}(t,x)$ is also a solution of Eq. (\ref{ev1}), then the system (\ref{ev1})
is said to be invariant under the scaling transformation (\ref{scaltr}). In
the following we denote, for simplicity, $y_{i}^{\left( s,\vec{\alpha},\beta
\right) }$ by $\stackrel{(L)}{y_{i}}$, and assume that generally $L>0$.

The renormalization group transformation $\Re _{s}$ is defined as a
transformation acting on the phase space $\Gamma $ of the system (\ref{ev1}) \cite{HaKoAd96}:
\begin{equation}
\Re _{L}:Y_{i}\left( \xi \right) =y_{i}\left( 1,\xi \right) \rightarrow 
\stackrel{(L)}{Y_{i}}\left( \xi \right) =\overset{(L)}{y_{i}}\left( 1,\xi
\right) =e^{\alpha _{i}L}y_{i}\left( e^{-L},e^{-\beta L}\xi \right) .
\label{rg}
\end{equation}

More exactly, $\overset{(L)}{Y}$ is given by developing the initial data $%
y\left( 1,x\right) \equiv Y\left( x\right) $ from $t=1$ to $t=e^{-L}$ by Eq.
(\ref{ev1}), and then rescaling the spatial coordinate $x$ and the unknown
functions $y_{i}$. Because of the invariance under scaling, $\Re _{L}$ forms
a semi-group with the property $\Re _{L_{1}+L_{2}}=\Re _{L_{2}}\circ \Re
_{L_{1}}$. From this property one can find the infinitesimal generator $\dot{%
\Re}$ of $\Re _{L}$, given by the derivative of $\Re _{L}$ with respect to $L
$ at $L=0$, $\dot{\Re}\equiv \lim_{L\rightarrow 0}\frac{\Re _{s}-1}{L}$,
leading to $\Re _{L}=\exp \left( L\dot{\Re}\right) $.

A fixed point of the renormalization group $\left\{ \Re _{L}\mid L\in
R\right\} $ is a point $Y^{\ast }$ in $\Gamma $ satisfying the equation $\Re
_{L}\left( Y^{\ast }\right) =Y^{\ast }$ for all $L>0$. Equivalently, it can
also be defined as $\dot{\Re}\left( Y^{\ast }\right) =0$. A function $y$ is
called self-similar with parameters $\left( \vec{\alpha},\beta \right) $ if
it satisfies $y\left( t,x\right) =y^{\left( L,\vec{\alpha},\beta \right)
}(t,x)$. Each self-similar solution $y_{ss}$ of Eq. (\ref{ev1}) with
parameters $\left( \vec{\alpha},\beta \right) $ is related to a fixed point $%
Y^{\ast }$ of $\Re _{L}$ by $y_{ss}\left( t,x\right) =\left( -t\right)
^{\alpha }Y^{\ast }\left[ x\left( -t\right) ^{-\beta }\right] $ \cite{HaKoAd96}.

As a next step we introduce the tangent map of $\Re _{L}$ at $Y$. It is
defined by $T_{L,Y}\equiv \lim_{\epsilon \rightarrow 0}\frac{\Re _{L}\left(
Y+\epsilon K\right) -\Re _{L}\left( Y\right) }{\epsilon }$. For a fixed $L$
and $Y$,  $T_{L,Y}$ is a linear operator, related to the renormalization
group by the following  formal relation: $\Re _{L}\left( Y+K\right) =\Re
_{L}+T_{L,Y}\left( K\right) +O\left( K\right) ^{2}+...$. At a fixed point $%
Y^{\ast }$ of $\Re _{L}$, the tangent map is $T_{L,Y^{\ast }}=T_{L}\left(
K\right) \equiv \lim_{\epsilon \rightarrow 0}\frac{\Re _{L}\left( Y^{\ast
}+\epsilon K\right) -\Re _{L}\left( Y^{\ast }\right) }{\epsilon }$. The
tangent map at the fixed point forms a semi-group. In terms of the
infinitesimal generator $\dot{T}=\lim_{L\rightarrow 0}\frac{T_{L}-1}{L}$,
the tangent map can be represented as $T_{L}=\exp \left( L\dot{T}\right) $ \cite{HaKoAd96}. 

The flow of the renormalization group transformation near the fixed point $%
Y^{\ast }$ is determined by the eigenmodes of $\dot{T}$, with eigenvalues $%
k\in \mathbb{R},\mathbb{C}$, defined as $Y^{\ast }K=kK$. A mode with $\operatorname{Re}k>0$ is called
a relevant mode, and it is tangent to a flow diverging from $Y^{\ast }$. A
mode with $\operatorname{Re}k<0$ is tangent to a flow converging to the fixed point.
Modes with $\operatorname{Re}k=0$ are called marginal. If there is a finite number
of relevant modes, or no relevant mode, then the long-time behavior of
nonlinear partial equations can be predicted from the study of the
perturbations around the fixed point. 

To make the abstract approach presented above more clearer from the point of
view of the physical applications, we discuss the RG method for an equation
of the type \cite{BrKuLi94}
\begin{equation}
\dot{y}=y^{\prime \prime }+F\left( y,y^{\prime },y^{\prime \prime }\right) ,
\label{example}
\end{equation}
where a dot and a prime denote the derivatives with respect to the variable $%
t$ and $x$, respectively. We are interested in the asymptotics of the
solution of the form $y\left( t,x\right) \sim t^{-\alpha /2}f\left(
t^{-1/2}x\right) $, as $t\rightarrow \infty $. The RG methods transforms the
problem of large time limit into an iteration of a fixed time problem,
followed by a scaling transformation. The scale-invariant solution emerges
then as a fixed point of a map in the space of initial data, the RG map, and
the stability analysis becomes the analysis of the stability of the fixed
point under the RG \cite{BrKuLi94}. 

Let us fix for the Eq. (\ref{example}) some (Banach) space of initial data $%
\Sigma $. Next, we chose a number $L>1$ and set $\overset{(L)}{y}\left(
t,x\right) =L^{\alpha }y\left( L^{2}t,Lx\right) $, where $y$ solves 
Eq. (\ref{example}) with the initial data $f\in \Sigma $. The RG map $\Re
:\Sigma \rightarrow \Sigma $ is in this case $\left( \Re f\right) \left(
x\right) =\overset{(L)}{y}\left( 1,x\right) $. $\overset{(L)}{y}\left(
t,x\right) $ satisfies the equation 
\begin{equation}
\frac{\partial \overset{(L)}{y}\left( t,x\right) }{\partial t}=\frac{%
\partial ^{2}\overset{(L)}{y}\left( t,x\right) }{\partial x^{2}}+
\overset{(L)}{F}\left( \overset{(L)}{y}\left( t,x\right) ,\frac{\partial \overset{(L)}{y}\left( t,x\right) }{\partial x},\frac{\partial ^{2}\overset{(L)}{y}%
\left( t,x\right) }{\partial x^{2}}\right) ,
\end{equation}
where  $\overset{(L)}{F}\left( a,b,c\right) =L^{2+\alpha }F\left(
L^{-\alpha }a,L^{-1-\alpha }b,L^{-2-\alpha }c\right) $. The renormalization
group transformation depends on $L$ and the functional form of $F$. The
semi-group property of the transformation implies $\Re _{L^{n},F}=\Re
_{L,F^{n-1}}\circ ...\circ \Re _{L,F_{L}}\circ \Re _{L,F}$. Letting $t=L^{2n}
$, the long time asymptotics of the solution of  Eq. (\ref{example}) \ can
be obtained by iterating the fixed time problem: $y\left( t,x\right)
=t^{-\alpha /2}\left( \Re _{L^{n},F}f\right) \left( t^{-1/2}x\right) $, for $%
t\rightarrow \infty $. 

If there is an $\alpha $ so that we can find two functions $F^{\ast }$ and $%
f^{\ast }$ so that $F_{L^{n}}\rightarrow F^{\ast }$ and $\Re
_{L^{n},F}f\rightarrow f^{\ast }$, so that $F^{\ast }$ satisfies the scale
invariant equation $\dot{y}=y^{\prime \prime }+F^{\ast }$ and $f^{\ast }$ is
the fixed point corresponding to this equation, defined as $\Re _{L,F^{\ast
}}f^{\ast }=f^{\ast }$, then the asymptotics of the initial problem are
given by $y\left( t,t^{-1/2}x\right) \sim t^{-\alpha /2}f^{\ast }\left(
x\right) $. Therefore the long time behavior of the differential equation is
determined by the fixed point in the RG group, and the stability analysis
can be also performed by studying the perturbations around this point.

\section{Renormalization group transformations and the equations of state}

The energy momentum tensor of a bulk viscous cosmological fluid filling a flat
FRW Universe, with line element:
\begin{equation}\label{1}
ds^{2}=dt^{2}-a^{2}(t)\left(  dx^{2}+dy^{2}+dz^{2}\right),
\end{equation}
is given by \cite{Ma95, Ma96}:
\begin{equation}\label{2}
T_{i}^{k}=\left(  \rho+p+\Pi\right)  u_{i}u^{k}-\left(  p+\Pi\right)
\delta_{i}^{k},
\end{equation}
where $\rho$ is the energy density, $p$ the thermodynamic pressure, $\Pi$ the
bulk viscous pressure and $u_{i}$ the four velocity satisfying the condition
$u_{i}u^{i}=1$. We use units so that $8\pi G=c=1$.

The gravitational field equations, together with the continuity equation
$T_{i;k}^{k}=0$, imply
\begin{eqnarray}
3H^{2}=&&\rho, \label{fe1}\\
2\dot{H}+3H^{2}=&&-p-\Pi,\label{3}\\%
\dot{\rho}+3\left(  \rho+p\right)  H=&&-3H\Pi,\label{4}%
\end{eqnarray}
where $H=\dot{a}/a$ is the Hubble parameter.

The causal evolution equation for the bulk viscous pressure is given by
\cite{Ma95, Ma96}
\begin{equation}
\tau\dot{\Pi}+\Pi=-3\xi H-\frac{1}{2}\tau\Pi\left(  3H+\frac{\dot{\tau}}{\tau
}-\frac{\dot{\xi}}{\xi}-\frac{\dot{T}}{T}\right) ,\label{5}
\end{equation}
where $T$ is the temperature, $\xi$ the bulk viscosity coefficient and $\tau$
the relaxation time. Eq. (\ref{5}) arises as the simplest way (linear in $\Pi
$) to satisfy the $H$ theorem ( i.e., for the entropy production to be
non-negative, $S_{;i}^{i}=\Pi^{2}/\xi T\geq0$, where $S^{i}=eN^{i}-\frac
{\tau\Pi^{2}}{2\xi T}u^{i}$ is the entropy flow vector, $e$ is the entropy per
particle and $N^{i}=nu^{i}$ is the particle flow vector) \cite{IsSt76,HiLi89}.
When the the condition $\frac{T}{a^{3}H}\left| \Pi \frac{d}{dt}\left( \frac{%
a^{3}\tau }{\xi T}\right) \right| <<1$ holds, the additional terms to the
evolution equation are negligible in comparison with $3\xi H$, and the
full equation leads to a truncated equation with reduced relaxation time
and reduced bulk viscosity \cite{Ma96}.

We also suppose that the bulk viscous cosmological fluid obeys the
barotropic equation of state
\begin{equation}
p=(\gamma-1)\rho, \qquad 1\leq\gamma\leq2.\label{6}
\end{equation}

With this choice the conservation equation (\ref{4}) can be formally integrated to
give
\begin{equation}
\rho(t)=\frac{1}{a^{3\gamma}}\left(  \rho_{0}+\int a^{3\gamma-1}\dot{a}\Pi
dt\right)  , \label{7}
\end{equation}
where $\rho_{0}\geq0$ is a constant of integration.

We consider the following scale transformations:
\begin{align}
t  & \rightarrow Lt,\label{8}\\
a(t)  & \rightarrow\overset{(L)}{a}(t)\equiv L^{n}a(Lt),\label{8.1}%
\end{align}
where $L>1$ is the parameter of the scale transformation and $n$ is a
constant. With the use of Eqs. (\ref{8})-(\ref{8.1}), from the Hubble parameter
definition and from the field equation (\ref{3}), it follows that the Hubble
parameter and the energy density are scaled in the following way:
\begin{align}
H(t)  & \rightarrow\overset{(L)}{H}(t)=LH(Lt),\label{9}\\
\rho\left(  t\right)   & \rightarrow\overset{(L)}{\rho}(t)=L^{2}\rho\left(
Lt\right)  .\label{9.1}%
\end{align}

From equations (\ref{3}) and (\ref{4}) we can now easily find the scaling
properties of the bulk viscous pressure and of the scale factor respectively:
\begin{align}
\Pi(t)  & \rightarrow\overset{(L)}{\Pi}(t)=L^{2}\Pi\left(  Lt\right)
,\label{10}\\
a(t)  & \rightarrow\overset{(L)}{a}(t)=L^{-\frac{2}{3\gamma}}%
a(Lt).\label{10.1}%
\end{align}

Eq.(\ref{5}) is scale invariant with respect to transformations (\ref{8})-(\ref{10.1}), if the
relaxation time $\tau$, the bulk viscosity coefficient $\xi$ and the
temperature $T$ obey the following scaling laws:
\begin{align}
\tau(t)  & \rightarrow\overset{(L)}{\tau}(t)=L^{-1}\tau\left(  Lt\right)
,\label{11}\\
\xi\left(  t\right)   & \rightarrow\overset{(L)}{\xi}(t)=L\xi\left(
Lt\right)  ,\label{11.1}\\
T(t)  & \rightarrow\overset{(L)}{T}(t)=L^{m}T\left(  Lt\right)  .\label{11.2}%
\end{align}
where $m$ is a constant. The causal evolution equation does not impose a
specific scaling law for the temperature.

In order to close the system of field equations (\ref{fe1})-(\ref{5}) we have
not only to give the equation of state for $p$, but also specify $T$, $\tau$
and $\xi$. There are many proposals in the physical literature for equations
of state for the relaxation time, temperature and bulk viscosity coefficient
\cite{HiSa91}, \cite{La82,BeNiKh79}.

Murphy \cite{Mu73}, Belinskii and Khalatnikov \cite{BeKh75} and Belinskii,
Nikomarov and Khalatnikov \cite{BeNiKh79} have analyzed the relativistic
kinetic equation for some simple cases. Their results show that in the
asymptotic regions of small and large values of density the viscosity
coefficients can be approximated by power functions of the energy density, with
definite requirements on the exponents of these functions. For small values of
the energy density it is reasonable to consider large exponents, equal, in the
extreme case, to one. For large $\rho$, the power of the bulk viscosity
coefficient should be considered smaller (or equal) to $\frac{1}{2} $. Hence,  in the following
we assume the following simple phenomenological laws \cite{Ma95, Ma96}, 
\cite{BeKh75, BeNiKh79}:
\begin{align}
\xi & =\alpha\rho^{s},\label{12}\\
T  & =\beta\rho^{r},\label{12.1}\\
\tau & =\frac{\xi }{\left( \rho +p\right) c_{b}^{2}}=\alpha\rho^{s-1},\label{12.2}%
\end{align}
where $c_{b}$ is the speed of bulk viscous perturbations, i. e., the
non-adiabatic contribution to the speed of sound in a dissipative fluid without heat flux
or shear viscosity \cite{Ma96},  and $\alpha\geq0$, $\beta\geq0$, $r\geq0$ and $s\geq0$ are constants.
Equations (\ref{12})-(\ref{12.2}) are standard in cosmological models, also
giving a simple procedure to ensure that the speed of viscous pulses does not
exceed the speed of light \cite{BeKh75, BeNiKh79}. With this choice we
obtain the following scaling equation for the bulk viscosity coefficient:
\begin{align}
\xi(t)  & =\alpha\rho^{s}\rightarrow\overset{(L)}{\xi}(t)=\alpha L^{2s}%
\rho^{s}\left(  Lt\right), \nonumber\\
\overset{(L)}{\xi}(t)  & =L^{2s}\xi\left(  Lt\right)  =L\xi\left(  Lt\right)
,\label{13}%
\end{align}
where we have also used Eqs. (\ref{11})-(\ref{11.2}). From Eq.(\ref{13}) it
immediately follows that a scale invariant bulk viscosity coefficient must
have $s=\frac{1}{2}$, and therefore the only physically acceptable behavior of
the bulk viscosity coefficient is of the form $\xi\sim\rho^{1/2}\sim H$ . In
fact, this result immediately follows from Eq. (\ref{13}), since the bulk
viscosity coefficient has the same scaling transformation law as the Hubble
parameter. In this way, in scale-invariant bulk cosmological models the bulk
viscosity coefficient is also determined by the expansion rate of the Universe. On the other
hand, $s=\frac{1}{2}$ is a case for which the gravitational field
equations for a flat causal bulk viscous cosmological fluid have been
intensively investigated in the 
physical literature \cite{PaBaJo91}, \cite{ChJa97}, \cite{MaHa98a,MaHa99a}.

The scaling property of the relaxation time can also be obtained easily and is
given by:
\begin{align}
\tau\left(  t\right)   & =\alpha\rho^{s-1}\rightarrow\overset{(L)}{\tau
}(t)=\alpha L^{2s-2}\rho^{s-1}(Lt),\nonumber\\
\overset{(L)}{\tau}(t)  & =L^{2s-2}\tau\left(  Lt\right)  =L^{-1}\tau\left(
Lt\right).\label{14}%
\end{align}

The scaling law (\ref{14}) holds again only for $s=\frac{1}{2}$.

In the context of irreversible thermodynamics $p$, $\rho$, $T$ and the
particle number density $n$ are equilibrium magnitudes which are related by
equations of state of the form $\rho=\rho\left(  T,n\right)  $ and $p=p\left(
T,n\right)  $. From the requirement that the entropy is a state function
\cite{Ma95}, we obtain the equation $\left(  \frac{\partial\rho}{\partial
n}\right)  _{T}=\frac{\rho+p}{n}-\frac{T}{n}\left(  \frac{\partial p}{\partial
T}\right)  _{n}$ . For the equations of state (\ref{12})-(\ref{12.2}), this
relation imposes the constraint $r=\left(  \gamma-1\right)  /\gamma$
\cite{Ma95}, so that $0\leq r\leq1/2$ for $1\leq\gamma\leq2$, a range of values
which is usually considered in the physical literature \cite{ChJa97}.
Therefore we obtain the following scaling law for the temperature of the
causal bulk viscous cosmological fluid obeying the state equations
(\ref{12}-\ref{12.2})
\begin{align}
T(t)  & =\beta\rho^{\frac{\gamma-1}{\gamma}}\rightarrow\overset{(L)}%
{T}(t)=\beta L^{2\frac{\gamma-1}{\gamma}}\rho^{\frac{\gamma-1}{\gamma}}\left(
Lt\right),  \nonumber\\
\overset{(L)}{T}(t)  & =L^{2\frac{\gamma-1}{\gamma}}T\left(  Lt\right)
.\label{15}%
\end{align}

An alternative set of equations of state has been used by Coley, van den
Hoogen and Maartens \cite{CoHoMa96} to study the full causal dissipative
evolution of a FRW space-time. The equations of state proposed in \cite{CoHoMa96}
are:
\begin{align}
\frac{\xi}{H}  & =3\xi_{0}x^{m},\label{16}\\
\tau^{-1}  & =Hx^{n},\label{16.1}\\
x  & \equiv\frac{\rho}{3H^{2}},\label{16.2}%
\end{align}
with $\xi_{0}$, $m$ and $n$ constants. By writing, with the use of Eqs.
(\ref{16})-(\ref{16.2}), the Einstein field equations as a plane autonomous
system, the qualitative behavior and the evolution of the Universe can be determined.

The scaling behaviors of the bulk viscosity coefficient and of the relaxation
time are given, for the phenomenological choice of the equation of state Eqs.
(\ref{16})-(\ref{16.2}), by
\begin{align}
\xi\left(  t\right)    & \rightarrow\overset{(L)}{\xi}\left(  t\right)
=3\xi_{0}LH\left(  Lt\right)  \frac{L^{2m}\rho^{m}\left(  Lt\right)  }%
{3^{m}L^{2m}H^{2m}\left(  Lt\right)  }=L\xi\left(  Lt\right)  ,\label{17}
\end{align}%
\begin{align}
\tau^{-1}\left(  t\right)    & \rightarrow\overset{\left(  L\right)  }{\tau^{-1}}=LH(Lt)\frac{L^{2n}\rho^{n}\left(  Lt\right)  }%
{3^{n}L^{2n}H^{2n}\left(  Lt\right)  }=L\tau^{-1}\left(  Lt\right)
\label{18}.
\end{align}

Therefore the set of equations of state (\ref{16})-(\ref{16.2}) are scale invariant for all $m$
and $n$.

For a flat FRW space-time, filled with a perfect fluid obeying a $\gamma $%
-law barotropic equation of state, the Hubble parameter $H$ induces a
natural time scale via the transformation $t\rightarrow t_{H}=\frac{2}{%
3\gamma }\frac{1}{H}$. The field equations are invariant with respect to
this transformation, having the form $3\left( \frac{1}{a\left( t_{H}\right) }%
\frac{da\left( t_{H}\right) }{dt_{H}}\right) ^{2}=\rho $ and $2\frac{%
dH\left( t_{H}\right) }{dt_{H}}+3\gamma H^{2}\left( t_{H}\right) =0$.
Therefore the RG analysis of the field equations for the perfect fluid can
also be performed in this time variable, and the RG transformation in the
Hubble time scale $t_{H}\rightarrow Lt_{H}$ leads to equations which are
formally similar to those obtained by using the usual cosmological time $t$.
The introduction of the bulk viscous pressure generally breaks the
invariance of the gravitational field equations with respect to the Hubble
time. In this case, by assuming a reparametrization of the cosmological time
of the form $t\rightarrow t_{0}/H$, $t_{0}=$constant, the $00$ component of
the field equations is again invariant, but the $ii$, $i\neq 0$ components
take the form  $2\frac{dH\left( t_{H}\right) }{dt_{H}}\left( \frac{3\gamma
t_{0}}{2}+\frac{t_{0}}{2}\frac{\Pi }{H^{2}}\right) +3\gamma H^{2}\left(
t_{H}\right) =-\Pi \left( t_{H}\right) $. This equation is scale invariant
only if the bulk viscous pressure is proportional to the energy density of
the fluid: $\Pi \sim H^{2}=\Pi _{0}\rho $, $\Pi _{0}=$constant. Then the
numerical coefficient in the scaling law of the time is given by $%
t_{0}=2/\left( 3\gamma +\Pi _{0}\right) $. For all other functional forms of
the bulk viscous pressure, the Hubble parameter does not generate an
invariant time scale for the Einstein field equations.   

\section{Renormalization group equations and long time behavior analysis}

Letting $t=L$, from the renormalization group transformations (\ref{8})%
-(\ref{11}) we obtain \cite{IgHoKo98}:
\begin{align}
a(t)  & =t^{\frac{2}{3\gamma}}\overset{(t)}{a}(1),\label{19}\\
H(t)  & =t^{-1}\overset{(t)}{H}(1),\label{19.1}\\
\Pi\left(  t\right)    & =t^{-2}\overset{(t)}{\Pi}(1),\label{19.2}\\
\xi\left(  t\right)    & =t^{-1}\overset{(t)}{\xi}(1),\label{19.3}\\
\tau\left(  t\right)    & =t\overset{(t)}{\tau}(1).\label{19.4}%
\end{align}

To obtain the renormalization group equations we introduce the parameter of
the renormalization group transformations in the form $L=e^{\theta}$. The
infinitesimal transformations of $\overset{(L)}{a}$, $\overset{(L)}{H}$ and
$\overset{(L)}{\Pi}$, with respect to $\theta$, are given by:
\begin{align}
\frac{d\overset{(L)}{a}}{d\theta}  & =-\frac{2}{3\gamma}\overset{(L)}{a}%
+\frac{\partial\overset{(L)}{a}}{\partial t},\label{20}\\
\frac{d\overset{(L)}{H}}{d\theta}  & =\overset{(L)}{H}+\frac{\partial
\overset{(L)}{H}}{\partial t},\label{21}\\
\frac{d\overset{(L)}{\Pi}}{d\theta}  & =2\overset{(L)}{\Pi}+\frac
{\partial\overset{(L)}{\Pi}}{\partial t}.\label{22}%
\end{align}

With the use of the (scaled) gravitational field equations and of the
evolution equation of the bulk viscous pressure the renormalization group
equations (\ref{20})-(\ref{22}) can be written as:
\begin{align}
\frac{d\overset{(L)}{a}}{d\theta}  & =-\frac{2}{3\gamma}\overset{(L)}%
{a}+\overset{(L)}{a}\overset{(L)}{H},\label{23}\\
\frac{d\overset{(L)}{H}}{d\theta}  & =\overset{(L)}{H}+\left[  -\frac{3\gamma
}{2}\overset{(L)}{H^{2}}-\frac{1}{2}\overset{(L)}{\Pi}\right] \label{24}\\
\frac{d\overset{(L)}{\Pi}}{d\theta}  & =2\overset{(L)}{\Pi}+\left[
-\frac{\overset{(L)}{\Pi}}{\overset{(L)}{\tau}}-3\frac{\overset{(L)}{\xi}%
}{\overset{(L)}{\tau}}\overset{(L)}{H}
-\frac{1}{2}\overset{(L)}{\Pi}\left(  3\overset{(L)}{H}+\frac
{1}{\overset{(L)}{\tau}}\frac{\partial\overset{(L)}{\tau}}{\partial t}%
-\frac{1}{\overset{(L)}{\xi}}\frac{\partial\overset{(L)}{\xi}}{\partial
t}-\frac{1}{\overset{(L)}{T}}\frac{\partial\overset{(L)}{T}}{\partial
t}\right)  \right] \label{25}%
\end{align}

The fixed point $\left(  a^{\ast},H^{\ast},\Pi^{\ast},\xi^{\ast},\tau^{\ast
},T^{\ast}\right)  $ of the RG equations is given as a solution of the system
of algebraic equations obtained by setting \cite{IgHoKo98}
\begin{align}
\frac{d\overset{(L)}{a^{\ast}}}{d\theta} & =0,\frac{d\overset{(L)}{H^{\ast}}}{d\theta}=0, \frac{d\overset{(L)}{\Pi^{\ast}}}{d\theta} =0.\label{26}
\end{align}

From the flow in the RG around the fixed point we can see the long time-behavior of causal bulk-viscous fluid filled flat Universe, once the equations
of state for the bulk viscosity coefficient, relaxation time and temperature
are given.

As an application of the mathematical formalism presented above we shall
consider the long time-behavior of the causal bulk viscous cosmological model
obtained by adopting the equations of state (\ref{16})-(\ref{16.2}), with $s=\frac{1}{2}$
and $r=\frac{\gamma-1}{\gamma}$. In this case the gravitational and causal
bulk evolution equations are scale-invariant. The equations of state take the
simple form $\xi=\sqrt{3}\alpha H$, $\tau=\frac{\alpha}{\sqrt{3}H}$ and
$T=3^{r}\beta H^{2r}$. The causal bulk viscous pressure evolution equation is
\begin{equation}
\tau\dot{\Pi}+\Pi=-3\xi H-\frac{1}{2}\tau\Pi\left[  3H-2\left(  1+r\right)
\frac{\dot{H}}{H}\right], \label{27}%
\end{equation}
while the RG equation (\ref{25}) becomes
\begin{align}
\frac{d\overset{(L)}{\Pi}}{d\theta}  & =2\overset{(L)}{\Pi}+\left[
-\frac{\overset{(L)}{\Pi}}{\overset{(L)}{\tau}}-3\frac{\overset{(L)}{\xi}%
}{\overset{(L)}{\tau}}\overset{(L)}{H}
 -\frac{1}{2}\overset{(L)}{\Pi}\left(  6\gamma\overset{(L)}%
{H}+\left(  1+r\right)  \frac{\overset{(L)}{\Pi}}{\overset{(L)}{H}}\right)
\right]  .\label{28}%
\end{align}

The fixed point of the RG equations (\ref{23}),(\ref{24}) and (\ref{28}) is
defined by
\begin{align}
a^{\ast}  & =\left(  \frac{3\gamma^{2}\rho_{0}}{4}\right)  ^{1/3\gamma
}, H^{\ast} =\frac{2}{3\gamma}, \rho^{\ast} =\frac{4}{3\gamma^{2}},\label{29}\\
\Pi^{\ast} & =0,\xi^{\ast} =0, \tau^{\ast-1} =0 \label{29.5}.
\end{align}

The fixed point of the RG equations corresponds to a flat, non-viscous
Friedmann Universe. In order to study the flow in the RG around the fixed
point, we must study the perturbation around the fixed point. We define
the perturbed quantities $\delta\overset{(L)}{a}$, $\delta\overset{(L)}{H}$,
$\delta\overset{(L)}{\Pi}$, $\delta\overset{(L)}{\xi}$ and $\delta
\overset{(L)}{\tau^{-1}}$ by \cite{IgHoKo98}
\begin{align}
\overset{(L)}{a}  & =a^{\ast}+\delta\overset{(L)}{a},\label{30}\\
\overset{(L)}{H}  & =H^{\ast}+\delta\overset{(L)}{H},\label{30.1}\\
\overset{(L)}{\Pi}  & =\Pi^{\ast}+\delta\overset{(L)}{\Pi},\label{30.2}\\
\overset{(L)}{\xi}  & =\xi^{\ast}+\delta\overset{(L)}{\xi}=\sqrt{3}%
\alpha\delta\overset{(L)}{H},\label{31}\\
\overset{(L)}{\tau^{-1}}  & =\tau^{\ast-1}+\delta\overset{(L)}{\tau^{-1}%
}=\frac{\sqrt{3}}{\alpha}\delta\overset
{(L)}{H},\label{31.1}%
\end{align}
where the perturbations $\delta\overset{(L)}{a}$, $\delta\overset{(L)}{H}$,
$\delta\overset{(L)}{\Pi}$, $\delta\overset{(L)}{\xi}$ and $\delta
\overset{(L)}{\tau^{-1}}$ are assumed to be small quantities, $\delta
\overset{(L)}{a}<<1$ etc. From the RG equations (\ref{23}), (\ref{24}) and
(\ref{28}), it follows that the perturbed quantities satisfy the following
linearized differential equations
\begin{align}
\frac{d\delta\overset{(L)}{a}}{d\theta}  & =a^{\ast}\delta\overset{(L)}%
{H},\label{32}\\
\frac{d\delta\overset{(L)}{H}}{d\theta}  & =-\delta\overset{(L)}{H}-\frac
{1}{2}\delta\overset{(L)}{\Pi},\label{33}\\
\frac{d\delta\overset{(L)}{\Pi}}{d\theta}  & =-\frac{4}{\gamma^{2}}\delta\overset{(L)}%
{H}.\label{34}%
\end{align}

From Eqs. (\ref{32}) and (\ref{33}) we obtain
\begin{equation}
\frac{d^{2}\delta\overset{(L)}{a}}{d\theta^{2}}+\frac{d\delta\overset{(L)}{a}%
}{d\theta}+\frac{a^{\ast}}{2}\delta\overset{(L)}{\Pi}=0,\label{35}%
\end{equation}
while Eqs. (\ref{33}) and (\ref{34}) give the following evolution equation for
the perturbed bulk viscous pressure:
\begin{equation}
\frac{d^{2}\delta\overset{(L)}{\Pi}}{d\theta^{2}}+\frac{d\delta\overset{(L)}{\Pi}}{d\theta}-\frac
{2}{\gamma}\left(  \frac{1}{\sqrt{3}\alpha}-\frac{1}{\gamma}\right)
\delta\overset{(L)}{\Pi}=0.\label{36}%
\end{equation}

The general solution of Eq. (\ref{36}) is given by
\begin{equation}
\delta\overset{(L)}{\Pi}\left(  \theta\right)  =C_{1}e^{r_{1}\theta}%
+C_{2}e^{r_{2}\theta},\label{Tony1}
\end{equation}
with $C_{1}$, $C_{2}$ constants of integration and
\begin{equation}
r_{1,2}=\frac{-1\pm
\sqrt{1+\frac{8}{\gamma
}\left(  \frac{1}{\sqrt{3}\alpha}-\frac{1}{\gamma}\right)  }}{2}.\label{Tony}
\end{equation}

From (\ref{Tony}) it follows that $r_{2}<0,\forall\alpha,\gamma$ while the sign
of $r_{1}$ depends on the sign of the quantity $\left(  \frac{1}{\sqrt
{3}\alpha}-\frac{1}{\gamma}\right)  $. If $\left(  \frac{1}{\sqrt{3}\alpha
}-\frac{1}{\gamma}\right)  >0$, ( $\gamma>\sqrt{3}\alpha$ ) then $r_{1}>0$
while for $\gamma<\sqrt{3}\alpha$, $r_{1}<0$.

With the use of Eq. (\ref{Tony1}), equation (\ref{35}), describing the evolution
of the perturbation in RG of the scale factor of the causal bulk viscous fluid
filled Universe takes the form
\begin{equation}
\frac{d^{2}\delta\overset{(L)}{a}}{d\theta^{2}}+\frac{d\delta\overset{(L)}{a}%
}{d\theta}+\frac{a^{\ast}}{2}\left(  C_{1}e^{r_{1}\theta}+C_{2}e^{r_{2}\theta
}\right)  =0,
\end{equation}
and has the general solution given by
\begin{equation}
\delta\overset{(L)}{a}\left(  \theta\right)  =f_{1}e^{-\theta}+f_{2}%
e^{r_{1}\theta}+f_{3}e^{r_{2}\theta},\label{Tony2}
\end{equation}
where $f_{i},i=1,2,3$ are constants of integration.

From Eq. (\ref{Tony2}) we can see the flow in the RG around the fixed point.
A fixed point $y_{0}$ of a system $\dot{y}=F(y)$ is said to be stable if
for every neighbourhood $U$ of $y_{0}$, there is a neighbourhood $U^{\prime
}\leq U$ of $y_{0}$ such that every trajectory which passes through $%
U^{\prime }$ remains in $U$ as $t$ increases. The fixed point is
asymptotically stable if it is stable and there is a neighbourhood $U$ of $%
y_{0}$ such that every trajectory passing through $U$ approaches $y_{0}$ as $%
t$ tends to infinity. A fixed point which is not stable is said to be
unstable \cite{Wa98}.

For $\gamma<\sqrt{3}\alpha$ the fixed point is an attractor.
Then, by setting the initial profile in the vicinity of the fixed point (\ref{29})-(\ref{29.5}), the
flat causal bulk viscous Universe with bulk viscosity coefficient proportional
to the Hubble parameter will approach, in the long-time limit, the flat
Friedmann Universe with $a(t)=a^{\ast}t^{2/3\gamma}$. This point is also
asymptotically stable.
For $\gamma>\sqrt{3}\alpha$, there is a single relevant mode.
Therefore, if $\gamma>\sqrt
{3}\alpha$, the space-time will deviate from the flat Friedmann geometry since
there is a relevant mode $\delta a(t)=f_{2}t^{r_{1}}$. This mode corresponds
to a growing mode, and therefore in this case the fixed point is asymptotically
unstable.

The conditions of the asymptotic stability or instability of the solution
can also be formulated in terms of the speed of sound in the cosmological
fluid. Since $\sqrt{3\alpha }=\xi /H=\sqrt{3}\xi /\sqrt{\rho }=\sqrt{3\rho }%
\gamma c_{b}^{2}\tau $, the condition of asymptotic stability becomes $%
c_{b}^{2}>1/\sqrt{3\rho }\tau $. With the use of the relation $\tau =\alpha
\rho ^{-1/2}$, the conditions of asymptotic stability or instability are
given by $c_{b}^{2}>1/\sqrt{3}\alpha $ and $c_{b}^{2}<1/\sqrt{3}\alpha $,
respectively.    

\section{Discussions and final remarks}

In the present paper we have performed an analysis of the flat homogeneous
bulk viscous cosmological models by using the RG method, in the version developed for the
study of asymptotics of non-linear evolution equation. For the flat FRW
geometry the Einstein gravitational field equations are invariant with
respect to the scale transformations of the time, energy density and Hubble
parameter. Therefore it is required that the causal bulk
evolution equation must also be scale invariant with respect to the same
group of transformations,thus leaving invariant the full set of gravitational field
and bulk viscous pressure equations. This condition leads to specific scale transformation laws for
the bulk viscosity coefficient and relaxation time, and imposes strong
constraints on the equations of state relating these parameters to the
energy density of the cosmological fluid. By using the property of the scale invariance of the field equations, we have shown
that the most used system of equations of state, expressing the bulk
viscosity parameter and relaxation time as a power of the energy density, $%
\xi \sim \rho ^{s}$, $\tau \sim \rho ^{s-1}$, is scale invariant for only one
value of the exponent $s$. Of course this result crucially depends on the linear
barotropic equation of state adopted to describe the matter content of the very early Universe.
This equation of state is not satisfied in more realistic physical cases, in which, for example,  the
Maxwell-Boltzmann gas could provide a more appropriate description of the cosmological fluid \cite{HiSa91}.  

The RG analysis also provides a powerful method for the study of the
long-time asymptotics of the gravitational field equations. This can be done
by the study of the flow around the fixed point of the RG group equations.
For the cosmological model having the bulk viscosity coefficient
proportional to the Hubble parameter $H$, the fixed point corresponds to a
flat ideal (non-viscous) Friedmann type geometry. Due to the presence of
viscous effects the geometry of the perturbed Universe will deviate from
that of the fixed point, and the linear analysis of perturbations indicates
the final state of the bulk viscous Universe. In the considered cosmological
model the long-time behavior of the space- time depends on the ratio of the
constant $\alpha $ relating $\xi $ to $H$ and $\gamma $.

Bulk viscous cosmological models have been investigated in the past years
from many different points of view. The method of the renormalization group,
as applied in a systematic manner to this class of models, can become a
valuable tool for cosmologists leading to
a better understanding of the nature, behavior and properties of our
Universe.

\section*{Acknowledgements}

The authors would like to thank to the two anonymous referees, whose comments
helped to significantly improve the manuscript.

\end{document}